Title: **Untangling the Web of e-Research: Towards a Sociology of Online Knowledge**

Eric T. Meyer[1] & Ralph Schroeder[1]
[1]Oxford Internet Institute, University of Oxford, 1 St Giles, Oxford, OX1 3JS, United Kingdom
Tel: +44 (0) 1865 287210
Fax: +44 (0) 1865 287211
e-mails: eric.meyer@oii.ox.ac.uk & ralph.schroeder@oii.ox.ac.uk
Corresponding author: Eric Meyer

Abstract (168 words)
e-Research is a rapidly growing research area, both in terms of publications and in terms of funding. In this article we argue that it is necessary to reconceptualize the ways in which we seek to measure and understand e-Research by developing a sociology of knowledge based on our understanding of how science has been transformed historically and shifted into online forms. Next, we report data which allows the examination of e-Research through a variety of traces in order to begin to understand how the knowledge in the realm of e-Research has been and is being constructed. These data indicate that e-Research has had a variable impact in different fields of research. We argue that only an overall account of the scale and scope of e-Research within and between different fields makes it possible to identify the organizational coherence and diffuseness of e-Research in terms of its socio-technical networks, and thus to identify the contributions of e-Research to various research fronts in the online production of knowledge.

## 1 Introduction

e-Research[1] is a rapidly growing area in many fields of scholarship, from the natural sciences to the humanities, as research moves online and becomes increasingly distributed across larger-scale and multi-institutional collaborations. Such a shift will pose major challenges to the sociological understanding of science and technology, a field which has so far relied heavily on case studies of individual projects instead of, for example, comparing across cases with middle-range theories (Beaulieu, Scharnhorst, & Wouters, 2007). Other ways of understanding science, such as bibliometrics, scientometrics, and more recently webometrics (Thelwall, 2007) also face problems in the online world insofar as some outputs (such as the paper-only monograph in fields such as anthropology) may be invisible in terms of online traces, particularly when compared to the highly-cited online-only physics article. Finally, some new ways to gauge directions in science, such as mapping of fields in terms of their online presence (Shiffrin & Börner, 2004), have yet to be integrated within the sociology of science and organizational analyses of knowledge production.

Several essays for this special issue contribute to the rapidly growing area of analyzing online data about knowledge, and we also analyze e-Research with these type of data. In addition, we consider some of the factors which shape the online realm as such, which can be considered a more qualitative type of analysis, but also one that cuts across disciplinary boundaries (as do many quantitative analyses). We thus aim to help fill the gap between qualitative and quantitative approaches to understanding e-Research, as well as between different disciplinary approaches, including the sociology of science and technology, information science, and broader approaches to knowledge which include the humanities.

The aim of this article is to bridge these gaps by means of a fundamental reconceptualization. We will argue, first, that a sociology of *knowledge* is more appropriate to coping with e-Research than the sociology of *science and technology*. This is because the term 'knowledge' expands beyond science and technology to also include the humanities on the one hand, but it also encompasses a wider conception of technology than individual artefacts such as the infrastructures that provide repositories of data or texts. Second, as e-Research necessarily entails online networks, knowledge in this case is created by means of various configurations of scientific and technological tools and resources, which minimally include the components of software, electronic networks and the digital materials of scholarship. e-Research can thus be seen as a set of socio-technical relationships that are configured around networked projects and programmes with different capabilities for online knowledge production. This includes configurations such as EGEE (Enabling Grids for E-SciencE, arguably the single largest global e-Research effort) that cannot easily be identified as a single e-Research project, or as an e-Infrastructure, or as a national or international e-Research programme. Similarly, the TeraGrid initiative funded by the National Science Foundation is not so much a project or discipline-specific infrastructure, but it can rather be seen as a federated large technological system with a number of more and less integrated projects in a variety of disciplines. Third, we argue that areas like e-Research require a discipline-transcending novel approach that combines quantitative data analysis but also an account of the socio-technical forces which shape the online realm of knowledge.

[1] We subsume under 'e-Research' other labels such as 'e-Science', 'cyberinfrastructure', and 'e-Infrastructure'.

On this basis, we will argue for a novel approach to understanding science and scholarship in view of the increasing digitization of knowledge (Borgman, 2007). The domain of e-Research will be examined because of its complexity and emergent nature (Dutton & Jeffreys, forthcoming; Jankowski, forthcoming). Yet e-Research needs to be given a precise definition because otherwise this could be taken to include all research using a personal computer and an internet connection. Here we shall mean by e-Research the *use of digital tools and data for the distributed and collaborative production of knowledge*. This excludes, for example, the mere use of email for informal communication. Though of course email is part any scholarly communication, it has become ubiquitous, and thus is not part of our definition any more than is electricity. This begs the question of what constitutes a contribution to knowledge production – an issue we shall return to.

## 2 Sociology of Online Knowledge

[INSERT FIGURE 1 ABOUT HERE]

Figure 1 shows existing approaches to online knowledge and how the approach suggested here presents a more encompassing view. Existing approaches to understanding knowledge discovery, distribution and access are, by and large, contained within the disciplinary silos indicated here, with little crossover between and among the silos. For instance, the sociology of science and technology (SST, which includes various subdisciplines such as STS, SSS, and SSK[2]) has long been active in understanding the processes involved in the shaping of knowledge within science. However, the overlap between SST and Library & Information Science is very small, even though information scientists study many of the same communities by using different tools, including bibliometrics. Information Science does, however, overlap to some extent with the Scientometrics & Webometrics silo, which in turn has some overlaps with the Internet research silo. Researchers in these fields are, we would argue, aware of and reading each other's literature more than are those within the disconnected silos shown here. With regard to the disconnected silos, however, we argue that other than small numbers of interdisciplinary individuals who span and connect these disciplines, the majority of researchers are staying within their silos.

One way to bridge the gaps between these silos is by noting the objects of study they all share. The cloud of online knowledge depicted in Figure 1 is meant to represent the increasingly online world of knowledge, including not only publications, but also tools, data, and a variety of resources. This online knowledge includes both formal outlets, such as journals and data archives, as well as informal outlets, such as blogs, webpages, and podcasts. We use the cloud metaphor not to refer to 'cloud computing' initiatives, but rather to the amorphous nature of storage and access to these resources.

In the figure, the top and the bottom of the diagram represent the ends of a continuum, as indicated by the scale on the left, between materials that live purely online and those that are only available offline. The position of the disciplinary silos more toward the top or the bottom of the figure indicate the extent to which the objects of study in the discipline are mainly offline objects and behaviours, are mixed, or are mainly concerned with the online world. Thus Internet research is deeply engaged with the study of online phenomena, but also includes researchers doing work trying to understand the connection between the online and the offline. Yet some of these disciplines (the sociology of literature) are largely confined to offline knowledge (books and print journals), and even when they reach into the 'cloud' of online knowledge, they address only specific aspects of the shape of online knowledge. Other disciplines such as the examples of physics and chemistry are often still deeply engaged with offline tools and data, including their laboratory spaces, but are also increasingly engaged in understanding how the online world can enhance their work. So, while chemistry has traditionally focused on generating and using data, knowledge, tools, and so forth, and while sciences such as webometrics has mostly focused on studying the uses and structures of data, knowledge, tools, and so forth - there is overlap between them as well. The recent focus of many scientists on developing ontologies is one example of this (see, for instance, Catton, Sparks, & Shotton (2006)), while the physics' community's early adoption and active promotion of online distribution of pre-prints via arXiv.org as a model for the future of scientific communication is another (Ginsparg, 1996; Gunnarsdottir, 2005; Kling, 2004).

[2] Science & Technology Studies (STS), Social Studies of Science (SSS), and Sociology of Scientific Knowledge (SSK). For an overview of these approaches, see Hess (1997). Other useful points of departure into this literature include Hacket, Amsterdamska, Lynch, & Wajcman (2008) and journals such as *Social Studies of Science* and *Science, Technology and Human Values.*

Within the cloud of online knowledge, several trends are indicated. Increasing interdisciplinarity is a trend across many disciplines, particularly those involved in the area we focus on later in this article, e-Research. e-Research is by its very nature likely to be interdisciplinary, as various domain specialists, technology experts, and information specialists are routinely required to do the varied work required in e-Research. Tied to this is increasing collaboration, either between or among disciplines. Wuchty et. al. (2007) demonstrated an overall trend of increasing size of teams in the production of scientific articles over the second half of the 20$^{th}$ century, and our data presented below suggests that this is particularly acute with regard to e-Research. A third trend is the shift from specialized to generalized discovery of, distribution of, and access to knowledge. Rather than relying on subject librarians, subscriptions to paper journals, and specialized subject-specific databases, research is increasingly being discovered and accessed via general purpose tools such as Google and Google Scholar (Jacso, 2008). An important aspect of this increasingly generalized access is that regardless of the oft-heard complaint that search has killed the habit of browsing the library's collection of recent issues of subject journals; one is now more likely to discover research outside one's main discipline if topical search terms are used. For instance, a sociologist relying on browsing their library's sociology journals will be getting a well-distilled and, we would argue, narrow view of a topic such as the long-established field of network sociology. Relying on search, however, they are likely to at least be exposed to related material from other disciplines, such as the burgeoning study of on- and offline social networks within physics and other disciplines, and may also then choose to read and cite that material.

The structure of the cloud of online knowledge is also undergoing regular shaping and modification. Here we have indicated three forces influencing the shape: the technological infrastructures that are in place to support the online knowledge network, the structures supporting online publication (and the corresponding limitations including access restrictions and intellectual property control mechanisms), and the design and use of search tools. Thus, even if the trends indicated above tend towards the removal of barriers between disciplines, the affordances of the technologies making up the cloud can either lower or raise barriers, either by design or via unintended consequences of cumulative decisions.

Data-driven analysis of online knowledge, often based on the analysis of online networks or data establishing links between online publications, is bound to be incomplete because it lacks insights into these overall forces that shape online knowledge: the role that electronic infrastructures play in the dissemination of and access to knowledge. Elaborating this point for the whole of online knowledge is beyond the scope of this essay, but in the case of e-Research, this includes the technological infrastructures that are emerging, the physical networks that are connecting groups of researchers in different ways, and the channels for dissemination of and access to knowledge. These, in turn, are shaped by funding policies, different disciplinary practices, and the ways in which online publications are made available (e.g. via closed, limited, or open access).

The sociology of online knowledge, indicated in the figure by the dotted line encompassing aspects of many of the disciplines indicated here, is more inclusive than other disciplines which have addressed the workings of science and knowledge and the modalities in which they are conveyed. By cutting across the various disciplinary silos described above, the sociology of online knowledge avoids many of the biases of particular disciplinary approaches and extends across the whole realm of online knowledge.

The contributions in this special issue fit well with the argument we make here, although they tackle domains that are far more extensive than e-Research. Frenken et al. (2009) and Skupin (2009), for example, want to provide a spatial and geographical mapping of science; in other words, they want to map the shape of the ‘cloud’ of online knowledge (in their case, much wider than e-Research). Similarly, Chen et al. (2009) find ways to pinpoint particular breakthroughs within this cloud, just as Lambiotte and Panzarasa (2009) develop tools to identify communities within it. All these provide powerful quantitative insights into the shape of knowledge production and patterns. Our approach, in contrast, comes closest to that of Lucio-Arias and Leydesdorff, who include the systematicity of science communication and its relation to other parts of the social system in their analysis of publications. Although we do not consider this larger social system in our approach (though we do include the part of it related to e-Research), we go further than Lucio-Arias and Leydesdorff (2009) in not limiting ourselves to scientific publications, but also include the technological infrastructures and the technologies for data manipulation in our sociology of online knowledge (the focus on texts and documents, and the absence of technology, are well-known shortcomings of Luhmann’s sociology of science on which Lucio-Arias and Leydesdorff draw, and which Fuchs (2001), to be discussed shortly, has tried to remedy with his notion of the hard cores of technological networks). Thus we are able to provide a more comprehensive account of changes in knowledge production which include the technological channels and forces which underpin current changes in dissemination and access to knowledge.

## 3 Approaches to Transformations in Science

This is where it becomes all-important to further distinguish between knowledge in different fields or disciplines. Across all domains, e-Research entails manipulating knowledge; this can range from searching databases to identify gaps in life sciences research - to producing a searchable text of the works of a famous thinker in the case of humanities scholars. What needs to be acknowledged, however, is that the manipulation of knowledge takes more and less powerful forms, if we use as a yardstick of power the social and intellectual organization of the sciences in terms of uncertainty and interdependence (Whitley, 2000), or Hacking’s yardstick of representing and intervening in the physical world (Hacking, 1983), which can be applied to e-Research using the concept of ‘research technologies’ (Schroeder, 2008). On the organizational side, these yardsticks can be combined with Fuchs' idea of harder and softer network cores (Fuchs, 2001) around which knowledge networks congeal, harder in the case of natural sciences, and more diffuse in the humanities. In the humanities, a separate yardstick can also be used that revolves around the organization of 'schools' (Frickel & Gross, 2005) that form around certain networks and the competition within and between network structures (Collins, 1998). The social sciences are hybrids between the natural sciences and humanities, sometimes forming harder cores as when research technologies are used to replicate and extend data that can be manipulated, at other times developing schools around 'classical thinkers' or organizational cores around competing research programmes and schools of thought.

Such a sociology of knowledge, based on the organizational strength of socio-technical networks, must be underpinned by quantitative analysis. In section 5 below, we will use a combination of several sources of data which provide insights about the scholarly outputs related to e-Research, and regarding the overall resources devoted to different components of e-Research efforts. These data are not complete, but we will argue that any sociology of online knowledge such as the one we propose is bound to be provisional and should include not only quantitative data gathered from online sources (as the other contributions to this special issue do), but also attempt to understand the forces which shape the online realm of knowledge.

Before we do this, however, we must pursue the question of what constitutes a transformation in knowledge production? There are various ways to understand how science is changing. In this article, we limit the scope of this question to e-Research. To be sure, e-Research transforms knowledge in a various ways, but how? Some (for example, Hey & Trefethen, 2003) have claimed that a 'paradigm shift' or 'revolution' (as opposed to evolution) in knowledge is taking place, drawing on philosophical ideas about the nature of knowledge. Yet we will argue, first, that philosophy on its own is unlikely to be a useful guide to this transformation.

One way to understand e-Research might be to draw on Popper's philosophy, since the online and public nature of knowledge of e-Research meshes well with Popper's idea (lucidly explained by Magee, 1973, pp. 65-73) that only externalized knowledge can be regarded as valid knowledge. Popper proposed that 'objective knowledge' inhabits the public domain or what he called 'World Three' (World Three of the growth of human knowledge, for Popper, is distinct from World One, the physical world, and World Two, of individual mental states). Popper's main concern in arguing for valid and objective knowledge, however, was to demarcate science from non-science. In doing this, he took for granted, based on a premise about the evolution of knowledge throughout the course of human history, that science led to the growth of knowledge. And this evolutionary view of knowledge, in turn, also implicitly rests on the commonly held view that science precedes technological change.

Yet this evolutionary view of knowledge and of the relationship between science and technology has been shown to be at odds with what we know historically about science and its relationship to technology. Contra Popper, what has been established by comparative history is that instead of an evolutionary path, the growth of knowledge experienced a take-off at a particular period in history, shaped by the possibilities of an unfettered circulation of scientists in Europe, especially during the late 18$^{th}$ and early 19$^{th}$ centuries. Even more importantly, and again refuting the idea of an evolutionary growth of knowledge and that science drives technological progress, is that it was rather the other way around: research technologies drove science because they provided powerful tools generating 'high-consensus rapid-discovery' science (Collins, 1998, pp. 532-538; Schroeder, 2007, pp. 21-40). This take-off took place on the basis of research instruments during the 1600s. Still, we shall see that a more sociological notion of 'objective knowledge' or externalized knowledge, which focuses on the way in which knowledge is organized via more and less publicly accessible networks, or networks that are accessible throughout the relevant research communities, provides a useful way to think about why e-Research is driving knowledge advance.

A different way to understand e-Research is that it can be seen as a new paradigm in science and knowledge. Kuhn's idea of a paradigm has been invoked to make this claim (Kuhn, 1962). Yet, as is well known, Kuhn used this concept in different ways (Masterman, 1970). One interpretation of Kuhn's notion of a paradigm shift in science is simply as a fundamentally new way to do things. The other is that so much problematic evidence has accumulated that the old paradigm can no longer cope and a new paradigm (or framework) is needed.

Neither of these notions of paradigm fits the whole of e-Research. The first notion does not fit because there is no evidence of across-the-board shifts in science towards e-Research. Instead, it has been shown that there a major field differences (see, for example, Fry & Schroeder, forthcoming; Olson, Zimmerman, & Bos, 2008). The second would apply most readily to the problem of the 'data deluge' (Hey & Trefethen, 2003) as it has been argued that this problem requires a paradigm shift. Again, however, there are field differences (Borgman, 2007). Nevertheless, it is clear that there *are* urgent problems of data deluges in *some* disciplines. As we shall see, however, it is not just the data deluge per se that is the problem, but rather the technologies to cope with them.

Sociologists and historians have thus presented a much more historically accurate and sociologically realistic assessments of the growth of knowledge, based not on an inherent evolution of knowledge or revolutionary paradigm shifts, but on a specific account of when and why this growth occurs because of the research technologies and how they objectify knowledge in hard (as per Fuchs' ideas above) and replicable networks that generate consensus and allow science to move rapidly on to new territory. As we shall see, this conceptualization is directly relevant to e-Research, which revolves around tool development.

If not philosophy, what about information science? The scientometric perspective, based on a bibliometric analysis of citations and the half-life of citations, provide measures of which areas of science are most prominent or fast-moving. This perspective makes it possible to track the prominence of different fields historically, in terms of the rise and relative decline of the outputs in different disciplines or areas of research. This could also be done by means examining the inputs to research, which is an approach associated with the economics of innovation (Dosi, 1982; Freeman, 1984; Freeman & Soete, 1997) and typically measures R&D spending in different fields. In the case of e-Research, it would be necessary to examine not so much overall R&D spending, but rather the funding emphases within and between different government research programmes. As we shall see in the data below, both information science and the economics of innovation provide useful complementary approaches.

One missing element in understanding e-Research, in addition to research technologies, is how scientific 'communication' as a system (see also Lucio-Arias and Leydesdorff, this volume) has become the key to knowledge generation. Especially in the case of e-Research, this includes not only publications, but also getting access to whole sets of experimental results or other forms of data and research material. This kind of collaboration - sharing the tools, data and other materials that go into knowledge production - has become essential to *some* researchers. This is where mutual dependence and task certainty come in, because only in certain fields are these materials critical to progress. Put differently, wherever access to and being able to manipulate the materials is the bottleneck or reverse salient within the technological system (Hughes, 1983), the sciences depend on e-Research technologies to tackle these bottlenecks.

Networked tools and data can thus be seen as 'infrastructures' or 'large technological systems' (Hughes, 1994), although they are not 'infrastructures' in the conventional sense of supporting society-at-large since they are largely confined to a small community within society; namely, scientists and researchers. Note that there are two kinds of bottleneck: one within the discipline or field where the need for shared tools and data has become the bottleneck to making scientific progress, and the second is the way in which large technological systems or infrastructures have bottlenecks; namely, in the intertwining between technical and organizational components where there are typically interdependencies.

In any event, these infrastructures or large technological systems, consisting of online networks, are now integral to the functioning of certain areas within disciplines. *Where* they have become integral - in the sense that researchers have become mutually dependent on them and where task certainty revolves around them – that is the question that this essay must now seek to answer.

Before we do so, it can be noted that this mutual dependence and task certainty where e-Research technologies are needed to address bottlenecks – or technology pull – cannot be the sole factor in this spread since e-Research technologies are also being developed regardless of a need for them; in other words, the e-Research programmes are promoting a technology push. This is clear if we consider that many e-Research tools that are developed so far lack any discernible uptake.

It is also important not only to gauge to what extent different disciplines rely on e-Research technologies, but also *how* they do so. In other words, how integral is the organization of tool and data within a common infrastructure or large technological systems to the core of knowledge production in different fields? Fuchs (2001) has argued that only certain fields have networks that are organized surrounding cores of artefacts. If we combine an analysis of these networked artefactual and organizational cores (which can be measured, for example, by levels of Grid computing use) with a number of scientometric indicators and indicators of funding levels for research in different fields, then we will have a more complete picture of how e-Research is transforming knowledge (perhaps as complete and accurate as possible for a transformation that is still in the making).

In terms of quantitative measurement, it is in principle possible to obtain indicators of the total number of projects or size of funding or citations that relate to e-Research. However, since it can be assumed that the contours of e-Research will be homologous across a range of national efforts and publication outlets, it will be enough to provide a number of examples or illustrations and see if they point in the same direction.

The definition of e-Research includes ‘shared’ or ‘collaborative’ tools and data. These two terms should not be seen as being normative; instead, they simply indicate that e-Research as an enterprise by definition consist of tools and resources that more than one researcher can access or use. The fact that this sharing or collaborating takes place via networks is given. Data should be seen in the broad sense to include sources of information formalized for communication, interpretation and processing in a variety of disciplines (Borgman, 2007, pp. 119-120). And tools and data are interrelated in the sense that tools manipulate data - though here there are differences between sciences and humanities in this respect.

The approach that we put forward here, one that focuses on organizations and how their networks are shaped by research tools, must be complemented by taking into consideration the different practices of different disciplines. For example, the humanities are mainly digitizing texts, images, films, and audio recordings (Nentwich, 2003). Here it matters how organizations collaborate in the production of shared resources, rather than whether networked tools are used to manipulate text and images. Thus, in the case of humanities such as philosophy and literature, we can ask to what extent these materials are supported by or propagated within networks (Collins, 1998) in their aim of interpretation.

The transformation of knowledge by e-Research thus takes quite different forms in different domains. Inasmuch as the humanities are digitizing texts and other materials for shared use, the effect is primarily two-fold: tying researchers together around certain resources, and making texts and other materials amenable to digital manipulation. Neither is particularly new, though the effect of e-Research, as in the sciences, is to increasingly tie researchers together around shared data resources – their artefactual and organization network cores. This goes against the ‘lone’ scholar model that is prevalent in some humanities fields, and makes research more collaborative organizationally and technologically. The key question, however, is to what extent do various fields and disciplines depend on these large technological systems or infrastructures? The difficulty of answering this question in the case of the humanities is compounded by the habit of many humanities researchers to cite all materials, including digitized resources used exclusively online, as if the researcher had consulted the original paper resource. These conventions can make all traces of uses of shared digitized resources disappear in the published research outputs.

Finally, it could be argued that focusing on the size of project grants is misleading since in the humanities it is primarily individual scholarship that matters. However, this point merely highlights the limitations of humanities in e-Research, since individual scholars will have limited possibilities to maintain collaborative resources over the course of time without institutional and technical support.

## 4 Methods of Measuring e-Research

Considerable data are available to measure e-Research using a variety of lenses and viewpoints. As part of the Oxford e-Social Science project[3], we have collected data from a variety of sources, only some of which we have room to present here. At the level of national programmes in the US, UK, and EU, we have collected data on funding and number of projects as a proxy for the scale of resources devoted to e-Research. In the US, for example, according to the National Science Foundation Database[4], the Office of Cyberinfrastructure alone has funded 342 projects in the period 2000-2008 for a total of over $375M, distributed among 158 different organizations. Other organizations also have funded e-Research projects in the US, spread across any number of funding agencies. In the UK and EU, data is somewhat less transparently available, but it is possible to piece together data about the scale (numbers of projects and scale of funding) by drawing on a number of sources; Gentzsch (2006) reports that $400M was spent as part of the UK e-Science initiatives from 2001-2006 and involved up to 1100 scientists.

[3] Oxford e-Social Science project website: http://www.oii.ox.ac.uk/microsites/oess/

Such a breakdown brings us to the various socio-technical components of e-Research efforts, which includes tools, resources and infrastructures. One data source which can contribute to an understanding of the level of resources devoted to the different components of e-Research is the amount of shared computing power used by different disciplines over the course of time; for example, Atkinson (2006, p. 41) reports that engineering and the physical sciences somewhat unsurprisingly use the UK National Grid Service most heavily, while the humanities and social sciences are very light users. Data such as these support the typologies that distinguish between socio-technical components developed in different e-Research efforts; for example, between computation-centric, data-centric and community-centric efforts (David & Spence, 2003). Finally, by means of webometric analysis it can be shown how, nationally and internationally, e-Research efforts cluster around core projects, disciplines and national funding initiatives or infrastructures (Ackland, Fry, & Schroeder, 2007).

As indicated, while there are a number of ways to measure the impact of e-Research, here we must limit our discussion to several. First, we report on a bibliometric sample obtained from Scopus using a search string[5] designed to retrieve a wide variety of articles on e-Research. The search string was used to gather a wide set of articles by searching in all fields, and a subset that only considered appearance of the term with the title-abstract-keyword fields. This string yielded a total of 5776 articles from 1994-2008 in the wide set, including 2186 journal articles, 3020 conference papers and the remainder a variety of other academic outputs (primarily editorials and reviews). The narrower title-abstract-keyword set contains 1320 articles over the same 1994-2008 time period, including 451 journal articles and 692 conference papers. Scopus was chosen for this project for several reasons, including the availability of discipline and field categories available in the Scopus database. Scopus has also been found to be equivalent to the Web of Science in its coverage of journal publications, but superior in terms of conference presentations (Meho & Rogers, 2008; Meho & Yang, 2007), including in terms of conferences for the computer sciences that are central to many e-Research activities.

[4] National Science Foundation Database: http://www.nsf.gov/awardsearch/index.jsp
[5] Search term: ({e-Research} OR eResearch OR {e-Science} OR eScience OR {e-Humanities} OR eHumanities OR {e-social science} OR cyberscience OR {cyber-science} OR (cyber PRE/0 science) OR cyberinfrastructure OR (cyber PRE/0 infrastructure) OR {cyber-infrastructure} OR cyberresearch OR {cyber-research} OR (cyber PRE/0 research) OR {e-Infrastructure} OR eInfrastructure OR (humanities w/15 digital) OR ((humanities W/15 online) AND (humanities W/15 computing)) OR ("social science" w/5 infrastructure) OR ("social scientist" w/5 infrastructure) OR ("humanities" w/5 infrastructure) OR (grid w/5 humanities) OR (grid w/5 "social science"))
The search term includes multiple versions of spellings of words to capture as many variations as possible. So, for instance, {e-Research} in curly brackets maintains the hyphen in the exact search term, (cyber PRE/0 science) returns results with the word cyber immediately before the word science, (humanities w/15 digital) returns instances of the two terms in the same sentence, and (grid w/5 humanities) returns instances of the words used in the same phrase.

Second, we report on data pertaining to the projects funded by the US National Science Foundation through the Office of Cyberinfrastructure. Finally, we look at some data from Google Insights to understand the ways in which others are searching for terms related to e-Research. There are a number of other data sources available that we intend to report elsewhere, but for the purposes of this article, these are sufficient to illustrate some of the emergent structures underlying e-Research, particularly in the US and the UK.

All of these data sources are incomplete, but together they give a good indication of the focus and direction of the e-Research effort. Even the focus and direction, however, are only partial indicators of the changes in and impact on research practices. This is where it becomes important to look closely at the components of e-Research, and which part of the research process e-Research tools, data and materials are deployed in and integral to.

## 5 Results

The data reported here represent a lens through which we can start to see the patterns that are developing around e-Research in a variety of settings. Starting with the Scopus sample of articles on the general topic of e-Research (including a variety of associated search terms) we see a number of patterns that illustrate the development of publications in this area, disciplinary differences engaged in e-Research, and the language used in the titles of articles on this topic. We also include data from the National Science Foundation, partly discussed above but also included below to compare the language used in titles of grant proposals to those used in research outputs.

[INSERT FIG 2 ABOUT HERE]

In Figure 2, we can see the growth of articles on e-Research over time. In the figure, each set of bars is comprised of a left side and a right side. The data on the left half of the bar shows the total Scopus results for the search term indicated above searching in all fields in Scopus divided into two parts: the bottom portion indicates the number of articles using the narrower method of only searching for the term within the title, abstract and keywords, and the top portion shows the additional articles added by searching in all fields. The data on the right half of each bar indicates the percentage of the wider search comprised of articles, in the bottom portion, and conference proceedings, in the top portion. The total number of articles represented in this figure is 5206.

Except for a few scattered early articles, publications did not begin to appear in significant numbers until about 2003 and then saw rapid growth, particularly in 2004 and 2005. More recently, however, it is unclear whether the growth of publications is levelling off or possibly even started to shrink somewhat. The total number of articles in 2008 (n=1157) is about 4.6% less than 2007 (n=1213). It will be several years before it would be possible to say whether these numbers are a minor blip in the data or a trend. It is possible that some of the apparent downturn in the most recent data is due to sources that have not yet been updated in Scopus. Regardless of the overall trend, however, this simple figure does serve to illustrate the novelty of this newly developed area of research and funding. It is worth noting, however, that many of the actors engaged in this domain consider themselves to have been involved with e-infrastructure long before the actual term came into use: in the AVROSS sample reported by Barjak et al., over 10% of their sample reported having been engaged with e-infrastructure since before 1995 (Barjak et al., 2007, p. 24).

Also worth noting in Figure 2 is the change in the proportion of the sample comprised of journal articles and of conference papers. The majority of publications on e-Research in its nascent years were journal articles. From 2004, however, while the number of journal articles continues to grow steadily, much of the explosive growth in publications on e-Research is due to conference proceedings, with over half of the publications in the sample coming from conference proceedings. There is a fairly simple explanation for this, but it speaks to the sociology of how knowledge about e-Research is spreading: by 2004, conferences were being established specifically to discuss developments in e-Science and related areas. The UK e-Science All Hands meetings began in 2002 in Sheffield, UK, the first IEEE e-Science conference was held in 2005 in Melbourne, Australia, and the 1st International Conference on e-Social Science was held in Manchester, UK in 2005. These conferences generated publications, which appear online and began to influence the scholarly discourse, but also generated connections and future collaborations between the researchers who met at the conferences. These and other e-Research oriented conferences attract researchers from a variety of backgrounds, which weaken the disciplinary boundaries among e-Researchers and also, therefore, require bridging the disciplinary silos among those who study this type of online research.

[INSERT FIG 3 ABOUT HERE]

In Figure 3, the field differences among the manuscripts in the sample are reported. ‘Field’ differences are, of course, different from disciplinary differences, and we will use both depending on the data available below. Clearly, however, the two are related and differences in fields give good indications about disciplinary differences and vice versa. The fields used here are aggregated from the “Subject Area” data in Scopus. The subject area classifications fall into 30 categories, and are arrived at using a combination of human and computer agent classification (Griffiths, 2004). Of the 30 categories, 25 appeared in our sample. However, because many of them were quite small, we aggregated these into the nine broader fields reported here. Not all disciplines were collapsed into a field, however: physics, for instance, has been widely discussed in the literature on scholarly communication, and thus seemed to merit separate consideration from either the rest of the natural sciences. It was also kept separate from mathematics because of different rates of co-authorship in these fields reported elsewhere (Wuchty, Jones, & Uzzi, 2007).

In the bottom chart of Figure 3, the proportion of the sample represented by each of the fields is reported. Each field-specific bar has two portions: the bottom portion represents the percentage of articles within that field that are also identified as part of another field in the database, which is an indicator of multi-disciplinarity. As a result, the numbers in this figure add up to more than the total number of articles, since the multi-disciplinary articles are by definition represented multiple times, once in each field indicated. The top portion represents the percentage of articles that are only identified with a single field. Note here the field differences regarding multi-disciplinarity. Computer science is split 59% to 41% in favour of multidisciplinary articles, while engineering is split with the same percentages but in favour of single-field articles. Mathematics is nearly entirely multi-disciplinary in this sample (99%, n=882), but we suspect this may be an artefact of the Scopus methodology; it is possible that the category of mathematics was used for a wide variety of computational approaches to science, which in turn is represented well in an e-Research sample, rather than representing large numbers of mathematicians being involved in e-Research publications. When we examined the data for the articles coded within the mathematics subject area, a number of articles with which we were familiar did not contain contributions from mathematicians. This, of course, also suggests future work aimed at refining the categories we are using in this analysis beyond those provided by Scopus. In general, however, the numbers here are along the lines of what one would expect based on what has been reported elsewhere about collaboration within and among fields: medicine and physics are more likely to be multidisciplinary, the social sciences and other natural sciences somewhat less so.

The mean number of fields per article is a rough representation of the interdisciplinarity and collaborativeness of the articles in this sample. For these data, all the fields except the natural sciences are significantly different from the overall sample mean (1.52). This relatively low overall mean is due in part to the large number of single field articles in the largest fields in the sample, computer science and engineering. Nevertheless, the fields that stand out as being the most interdisciplinary, as measured by the number of fields associated with their publications, are mathematics (2.66), medicine/biomedicine/health (2.41), and computer science (1.89). These fields are to some extent the usual suspects in terms of collaborative publication across disciplines, but this also underscores the extent to which e-Research reflects, and potentially amplifies, field and disciplinary differences. Disciplines such as physics, which are already well understood to be heavily engaged in collaborative publication across disciplines, are also apparently particularly well-suited to become engaged in e-Research. While the data here are not sufficient to infer causality, we can speculate that there may be a system of positive feedback in which a tendency toward collaborative production of knowledge leads to a greater likelihood of involvement with e-Research, which in turn amplifies the likelihood of collaboratively producing knowledge using e-Research tools and data.

As has been reported elsewhere in data on all publications (Wuchty, Jones, & Uzzi, 2007) and discussed in more detail below, the mean number of authors per article varies considerably by field. In our sample of articles about e-Research, the overall mean number of authors is 3.67 authors per article, but this ranges from an average of 1.44 authors per article in the arts and humanities to 4.41 authors per article in physics. All of the fields except for engineering and mathematics are significantly different from the overall sample mean, as shown in the figure.

In terms of citation rates, all of the fields except for physics and the arts & humanities are significantly different from the overall sample mean. The number of citations ranges from a high of 6.22 in business & economics to a low of 1.83 in math; the overall average is 2.77 citations per article. The apparently high citation rate in business & economics is from a relatively small sample of articles, so not too much should be drawn from this, but see below for additional discussion on this point. In general, though, this portion of the figure gives a good indication of the citation practices and impact, in a bibliometric sense, of e-Research articles by field and discipline.

[INSERT TABLE 1 ABOUT HERE]

It is worth comparing the numbers in this sample for mean numbers of authors per article to that reported by Wuchty et al. (2007). As shown in Table 1, the mean number of authors in this sample is in general slightly higher than what Wuchty et al. report for their sample of all articles in the ISI database for the range of articles at the end of their sample period, 1996-2000. In their supplemental data, they report the mean values for size of teams, as indicated by authorship. Some of their categories do not match ours, and are not reported in a way that is possible to disaggregate, but we did run one-sample t-tests for those that are directly comparable. Of the 20 subject area classifications from our data which had directly comparable data in Wuchty et al., 15 were not significantly different in terms of the mean number of authors per paper (what Wuchty et al. refer to as “team size”). This gives us reason to believe that in general, our dataset is comparable to theirs in fields which are not generally seen to be central to e-Research (biochemistry or physics, for instance). However, in fields which are central to e-Research, as we will now show in more detail, there are significant differences in mean numbers of authors per paper.

So, in the case where the differences between their figures are ours are statistically significant, we may be seeing indications of a real difference in co-authorship for those involved with e-Research. The articles in the e-Research sample that have a higher mean number of authors include computer science (3.85 compared to 2.39), engineering (3.65 compared to 2.94), mathematics (3.85 compared to 1.84), business (2.52 compared to 1.66), and economics (2.25 compared to 1.71). At least the first three of these areas are central to many of the developments in e-Research. These differences in the size of their publication teams support the idea that their involvement in e-Research is also having an effect that can be measured in terms of their publication habits. For business and economics, the difference is a little less clear, but our speculation is that this may be to do with an apparent focus on network research in the business and economics sample, which may be more likely to be multiple authored than other work in those fields.

Caution is necessary, however. Without access to their complete dataset, it is difficult to know if differences across all fields are statistically significant. In addition, the end point of their dataset (1996-2000) corresponds with the earliest articles in our dataset, making it difficult to argue that the data is directly comparable. A comparison using better matched data, however, would be useful to pursue in a follow up study, since one argument made by the supporters of e-Research is that the tools and data enable larger collaborations. If the average number of authors per article publishing in the area of e-Research is higher than the average number of authors per article in the same fields publishing generally, this would support the notion. At this time, we can say we have a tentative indication that this is true, but need additional data to confirm or disconfirm this.

[INSERT FIG 4 ABOUT HERE]

This is illustrated even more strikingly in Figure 4. In this figure, the data we have been discussing has been converted to 3717 pairs of fields, as identified in the Scopus metadata. The diagram shows the comparative size of the number of articles in the collection associated with each field pairing with another field. Computer science clearly dominates, with 2463 paired fields in the sample, followed by mathematics (n=1541) and medicine/biomedicine/health (n=1358). More interestingly, the connections and sizes of connections (which of course are partly an artefact of the number of articles) give us an indication of which fields are collaborating on articles. The triumvirate of computer science, medicine, and mathematics are clearly quite interdependent. Computer science in particular is connected to nearly every field, except for the arts & humanities (which, again, is partly a reflection of the small sample size for this field, which resulted in no connections exceeding the minimum tie strength of 30 connections which was required to simplify the diagram). The social sciences are connected to computer science, more weakly to medicine, business & economics and engineering, but not at all to mathematics or physics. This indicates empirical support for the idea that that the so-called hard sciences are not actively collaborating with the social sciences and arts.

The dominance in the sample of computer science underscores the centrality of computer science to the entire e-Research enterprise. Since e-Research has focused on enabling sharing and collaboration of digital tools and data, computer science has understandably been a central player in e-Research. Even more than this, however, we would suggest based on other work (de la Flor & Meyer, 2008) that computer science has not just been the lucky recipient of the attentions of interested e-Researchers, but that many of the central actors pushing e-Research forward as a potential computerization movement (Elliott & Kraemer, 2008) are based in computer science. These actors have been actively engaged in efforts to enrol the participation of various domain researchers in e-Research. One open question, however, is whether this movement will succeed in the long term, and whether these computer scientists will be able to continue to expand computer science centric e-Research into other fields and disciplines, including those reported here.

One weakness of this sample is the comparatively small number of arts & humanities articles. There are several possible reasons for this. The most obvious is simply that the arts & humanities may be, by and large, not deeply engaged with e-Research. If this is true, further work may be needed to discover if this is due to active disengagement (i.e., rejecting potential involvement) or passive disengagement (i.e., lack of awareness of possibilities for involvement). A second potential reason was mentioned elsewhere in this article: the disciplinary habits and practices of arts & humanities scholars which tend to cause involvement with shared digital resources to disappear in the published outputs which favour citing resources, including electronic resources, as if the original paper sources had been consulted. A third reason has to do with the disconnect between preferred publication channels in the humanities, which favour book publication, and the content coverage in databases such as Scopus and ISI, which do not generally include books in their data. We are examining this in other ongoing research, but it is too early to report that work here.

[INSERT FIGURE 5 ABOUT HERE]

The last visualisation combines data from the Scopus sample with information from the US National Science Foundation Office of Cyberinfrastructure (NSF-OCI) in Figure 5. These word clouds are generated from all the words in the titles of the papers in the sample, with common English words removed. The top cloud is drawn from the Scopus data, the bottom cloud from the NSF-OCI database[6] of funded research, 2000-2008. This is a quick way to look at the topics in the articles and research grants based on the frequency of words in the titles. The figure also includes tables in the centre of the figure showing the frequencies of the top 25 words in each dataset. While these word clouds do not show any statistical relationship among various words in the titles, they do give an overall impression of the types of themes in our samples.

[6] http://www.nsf.gov/awardsearch/index.jsp

Several obvious foci are visible in the Scopus data: the focus on the Grid, data, applications, and computing are clear. Other common terms partly reflect the dominance of computer science in this sample: resources, workflow, service, computing, and semantic. The NSF-OCI data, however, provides a comparison that gives us a slightly different view on e-Research. The NSF-OCI has been an active and key player in advancing e-Research in the US through their funding efforts and around the world through their advocacy of cyberinfrastructure. Using the 341 projects in the database, the lower tag cloud represents at some level the ways in which scientists are selling their work to the funding agency. Comparing this to the tag cloud for publications, which represents at some level the results of these projects, makes clear some interesting differences. First, some concepts appear to be more likely to be discussed at the funding stage (collaborative, performance, demonstration) that are likely at least partly an effort to use the terms that the funders are using to discuss these relatively new efforts.

Note also that there is a focus on research, engineering and computing in both sets of data. Some relative frequencies are quite distinctive and worth noting. The term “collaborative” moves from being the fourth most frequent term in funding applications to being a relatively infrequent term in publications (visible in the cloud near the top). Computing remains important in both, but the Grid, only moderately frequently represented in funding applications, is absolutely central to the publications. Many other pairs of words lead us to speculate on the nature of transformation within projects over time. Undoubtedly, some of the difference here represents the inevitable difference between the type of language and strategic orienting that is required at the funding stage versus that required at the publication stage. Nevertheless, it makes us wonder to what extent the changes are also reflected in transformations within e-Research projects over time as initial expectations give way to operational reality.

## 5.1 Future work

This study is ongoing, as we collect additional data to help us understand the sociology of online knowledge, using e-Research as a case study. Future efforts will include additional analysis of the data reported here to include measures such as co-authorship networks and analysis of highly published and cited researchers. Examining authorship will also allow us to map more precisely the disciplinary affiliations of authors than the crude categories used within Scopus allow. We are also collecting webometric data using the same search terms we have used in the Scopus sample to find websites with these terms and build a dataset of the links between websites. This work will allow us to begin to map the global structure of online knowledge in the e-Research area.

# 6 Conclusion

In this article, we have discussed a number of ways to view the potential transformations of science promised by e-Research. By first discussing that we must start to develop a sociology of online knowledge if we are to understand this emergent domain, we have set out the position that understanding the socio-technical nature of e-Research is more complex than simply understanding the policy and funding programmes that have been set forth within the domain. To support this contention, we have presented data aimed at beginning to map the sociological features of this knowledge domain. At the same time, as we have seen, e-Research as a domain ranges across a number of disciplinary approaches that might analyze it – including information science, sociology of science, social studies of humanities, and studies of new media. As research increasingly moves online, within e-Research but also more broadly, the boundaries between these disciplinary approaches will need to be transcended in order to gauge the relations between fields and to map and identify the movements of the research fronts by means of online research tools, as the other contributions to this special issue do. This will not yield a complete picture, however, as these contributions also recognize: among the missing puzzle piece are structural characteristics such as the gatekeeping functions of search engines in shaping access to online research, the ways in which the publishing industry and publishing methods shape the balance between online and offline research resources, funding policies towards different e-Infrastructures, and the extent to which fields and disciplines use research technologies – tools, data and resources – with material that can be digitized.

The shaping of the online realm of knowledge has implications for the interpretation of Popper's notion of the visibility of research, as well as for potential Kuhnian paradigm shifts within certain fields. Whatever the differences between disciplines and fields, however, any sociology of online knowledge must recognize that research technologies are increasingly organized across institutional and disciplinary boundaries. The visibility and advance of the outputs of knowledge, apart from how it can be quantitatively measured, is also shaped by how these technologies provide a conduit for knowledge which can be analyzed as a separate online whole, even if there are continuities between online and offline knowledge.

The data in this article show that e-Research has taken hold in different fields of research to quite variable degrees. Much of this data suggests considerable future work to extend our understanding of this emergent domain, particularly research that combines quantitative and qualitative data. Neither purely quantitative nor qualitative approaches can do justice to e-Research for two reasons: one is that different sources of data need to be combined in order to obtain a middle-range and comprehensive understanding of the different levels of socio-technical networks involved. The second reason is that only an overall account of the scale and scope of e-Research within and between different fields makes it possible to identify the organizational coherence and diffuseness of e-Research in terms of its socio-technical networks, and thus to identify the contributions of e-Research to various research fronts in the online production of knowledge.

On the basis of what has been argued, we would expect the largest concentration of projects and the bulk of e-Research funding or resources to concentrate in those fields where the data deluge as conceptualized here (as mutually dependent communication) must be addressed as a matter of urgency. In certain fields more than others, there are bottlenecks in manipulating and organizing vast amounts of data and other materials by means of tools (Meyer, forthcoming). In other fields, such as the humanities, technological and organization interdependencies can be created even without these urgent needs or need for mutually dependent scholarship. How these research technologies form lasting network cores in knowledge production – and are able to build on and maintain them – is a further question which we have begun to answer, but which will need to be followed over the longer term.

What we can see already in the data is the emergence of concentrations of projects in certain disciplines – or rather in combinations of disciplines where computer science, which provides the research technologies, is connected to domain disciplines. We can also see higher levels of collaborative work around the use of research technologies across certain disciplines and fields as well as in higher levels of the use of shared computing power. If we come back now to the idea that objective knowledge is externalized and made visible around particular networks in technologically hardened cores with higher levels of task certainty and mutual dependence, we can see that these cores of research technologies - whether in the form of larger infrastructures or in projects that are not infrastructures in supporting whole communities of researchers but consist of smaller-scale socio-technical networks - these are more prominent in some disciplines and fields rather than others. How the technologies of e-Research will congeal still further and, if not revolutionize knowledge production then at least transform it and lead the edge of research in certain disciplines in new directions, remains to be seen. But certain network crystallizations are becoming visible already.

The approach here integrates data from a variety of sources and combines ideas about how knowledge changes with a sociological understanding of the material bases of these changes, via technological and socio-organizational networks. We have argued that only such a combination of sources and understanding of socio-technical networks can give an indication of the overall as well as the specific disciplinary directions of knowledge production in-the-making. Put differently, the sociology of knowledge needs to be based on the material organization of research technologies, which, in the case of e-Research, consists of an emerging landscape of socio-technical networks, the contours of which we have begun to sketch.

## 7 Acknowledgements

The work for this article has been supported by the Economic and Social Research Council (ESRC) grant RES-149-25-1022 and is part of the Oxford e-Social Science project (http://www.oii.ox.ac.uk/microsites/oess/). The authors are grateful to the editors of this issue, Katy Börner and Andrea Scharnhorst, and the anonymous reviewers for their helpful comments on this article.

## Figure Captions

Figure 1. Sociology of Online Knowledge

Figure 2. Publications on e-Research in Scopus, 1994-2008, (n=5206, articles and conference papers only)

Figure 3. Citation rates, authors, fields, field Size, and multidisciplinarity in Scopus e-Research articles, 1994-2008 (n=5206, articles and conference papers only)

Figure 4. Map showing number of articles by field, and article interdisciplinarity. The size of the nodes in relation to each other shows the comparative number of times each field was paired with another in the dataset. The line thickness connecting the nodes illustrates the number of connections between a given pairing. Pairs with fewer than 30 connections are not displayed in order to simplify the image.
Source: Data retrieved from Scopus using sample search terms; image created from data generated by Microsoft Excel NodeXL plugin.

Figure 5. Visualization of most common terms in titles of Scopus e-Research articles, 1994-2008 (n=5206) and most common terms in titles of NSF Office of Cyberinfrastructure grants awarded, 2000-2008 (n=341).
Source: Data from authors, Scopus (http://www.scopus.com), and the US National Science Foundation (http://www.nsf.gov/awardsearch/index.jsp); word cloud visualization and word counts from Wordle (http://wordle.net)

Figure 1

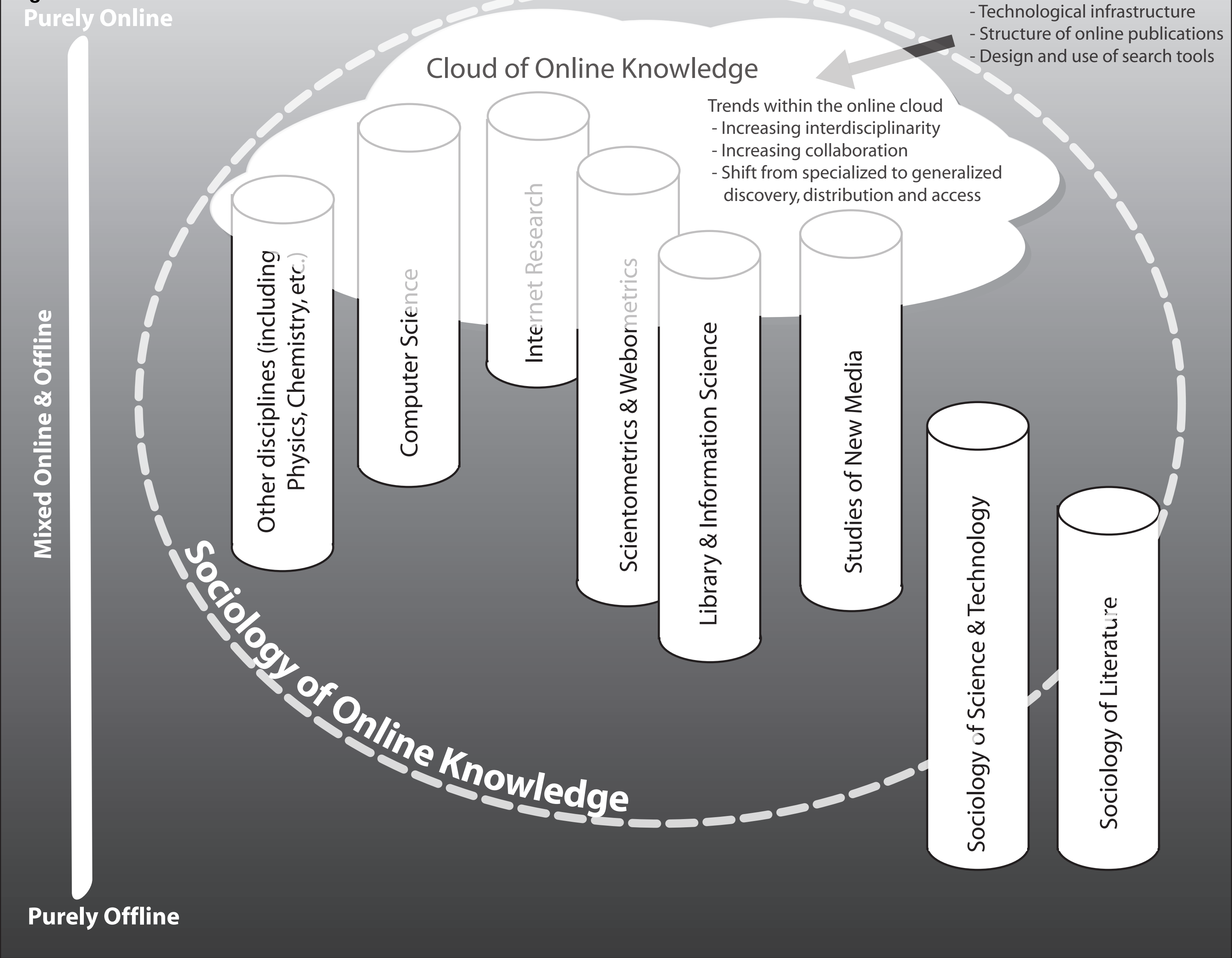
Purely Online
Mixed Online & Offline
Purely Offline
Cloud of Online Knowledge
Forces shaping the online cloud
- Technological infrastructure
- Structure of online publications
- Design and use of search tools
Trends within the online cloud
- Increasing interdisciplinarity
- Increasing collaboration
- Shift from specialized to generalized discovery, distribution and access
Other disciplines (including Physics, Chemistry, etc.)
Computer Science
Internet Research
Scientometrics & Webometrics
Library & Information Science
Studies of New Media
Sociology of Science & Technology
Sociology of Literature
Sociology of Online Knowledge

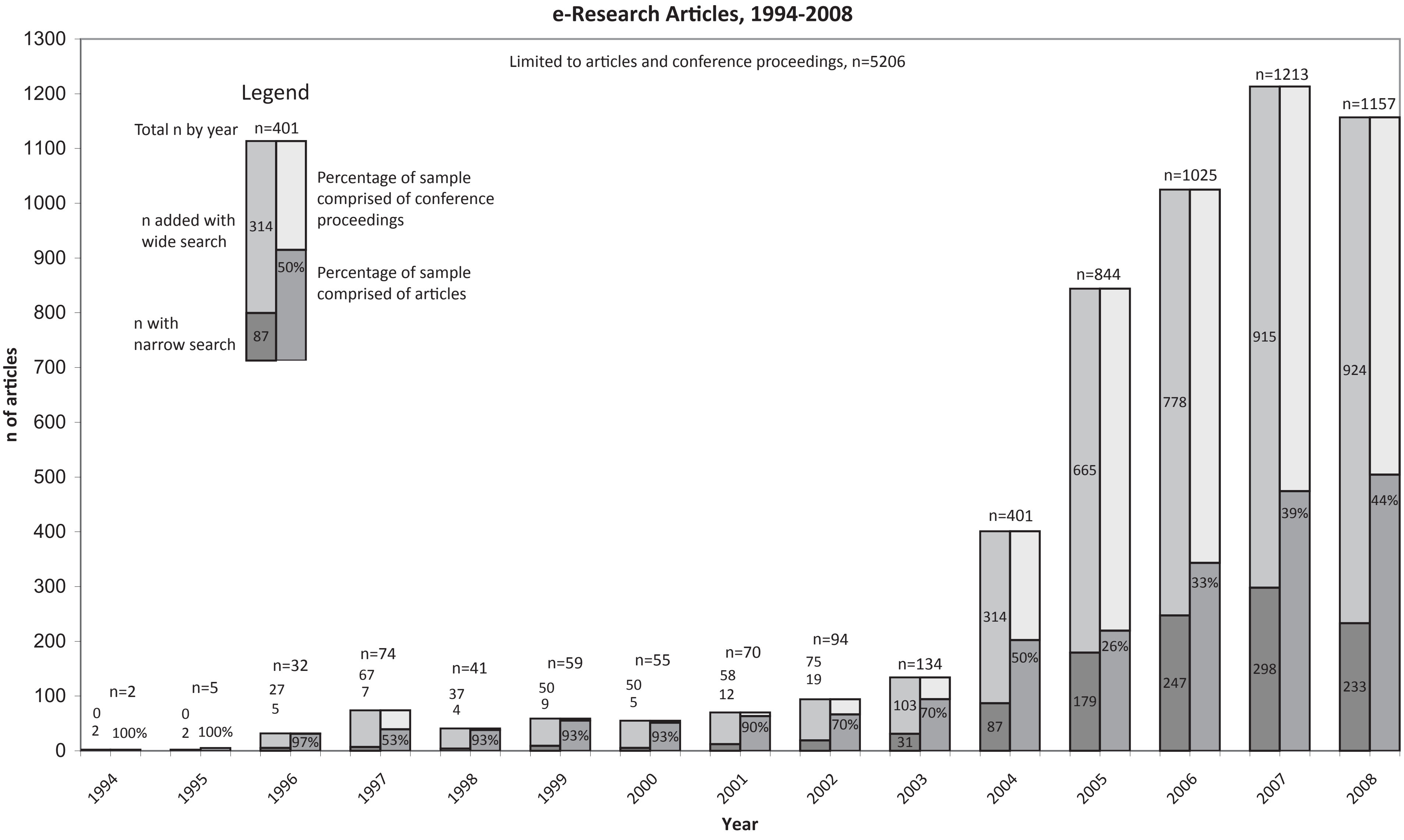
e-Research Articles, 1994-2008
Limited to articles and conference proceedings, n=5206
Legend
Total n by year
n=401
n added with wide search
314
n with narrow search
87
50%
Percentage of sample comprised of conference proceedings
Percentage of sample comprised of articles
n of articles
Year
1994 n=2 0 2 100%
1995 n=5 0 2 100%
1996 n=32 27 5 97%
1997 n=74 67 7 53%
1998 n=41 37 4 93%
1999 n=59 50 9 93%
2000 n=55 50 5 93%
2001 n=70 58 12 90%
2002 n=94 75 19 70%
2003 n=134 103 31 70%
2004 n=401 314 87 50%
2005 n=844 665 179 26%
2006 n=1025 778 247 33%
2007 n=1213 915 298 39%
2008 n=1157 924 233 44%

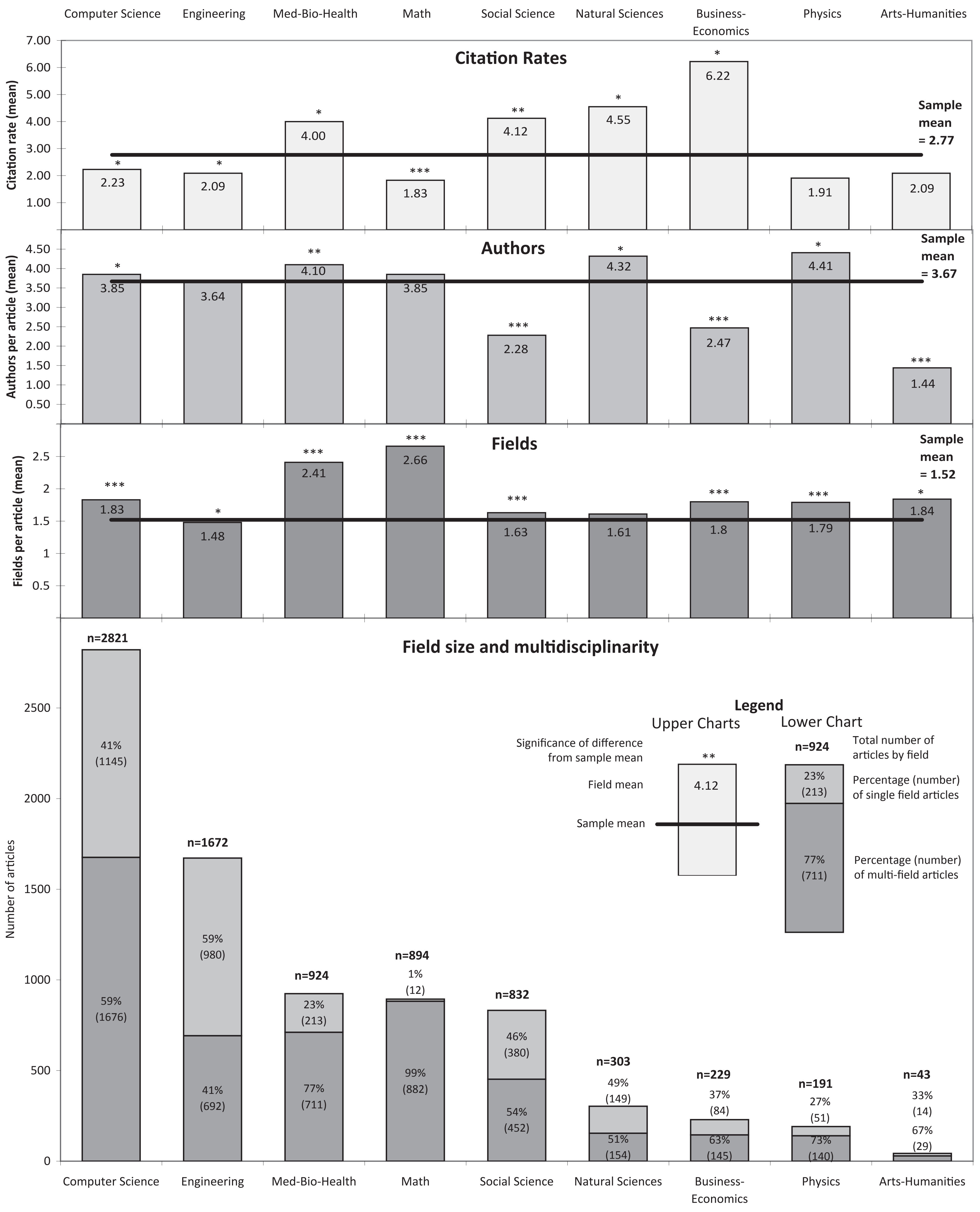

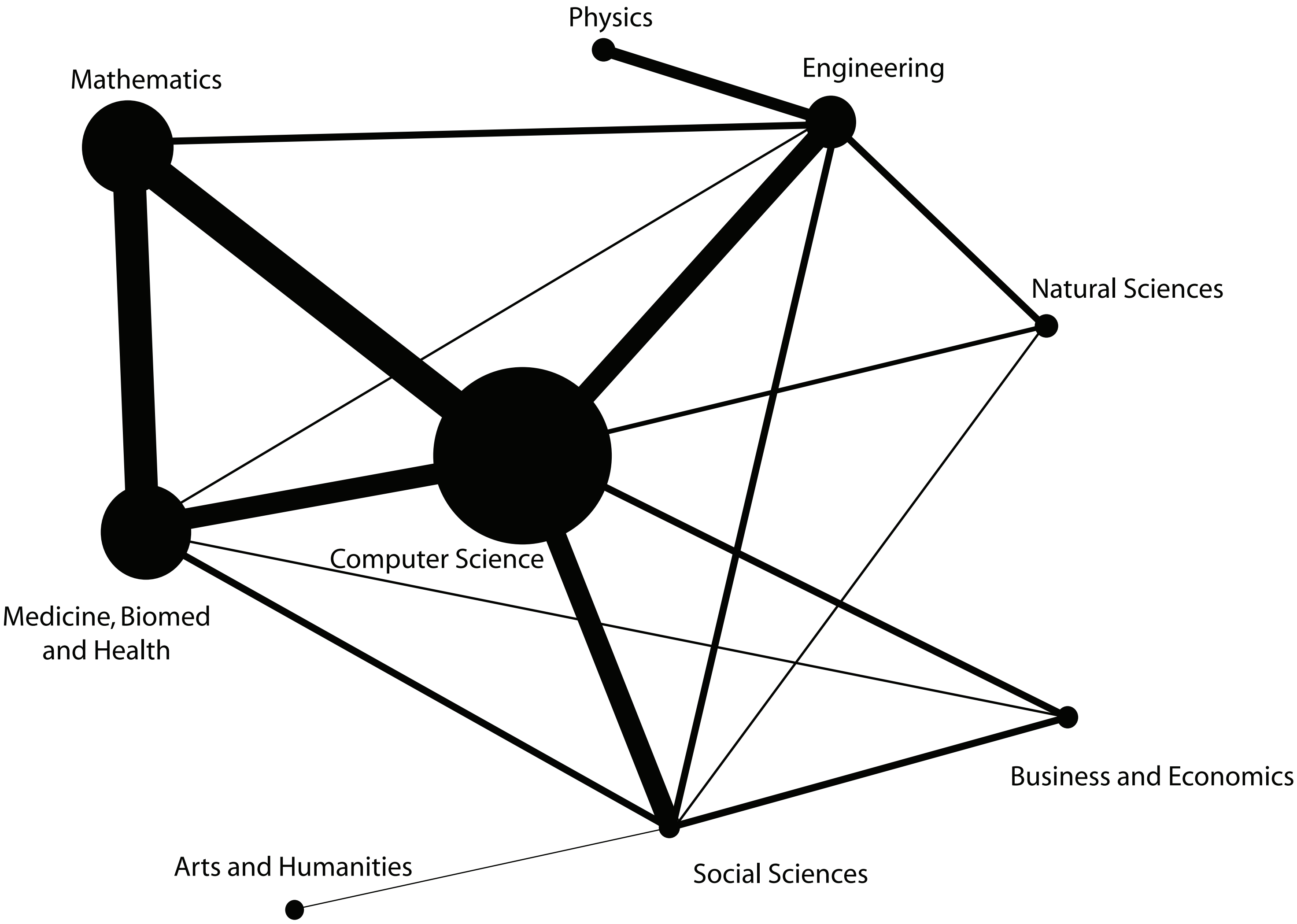

**Figure 5**

## Word cloud of Scopus e-Research titles, 1994-2008

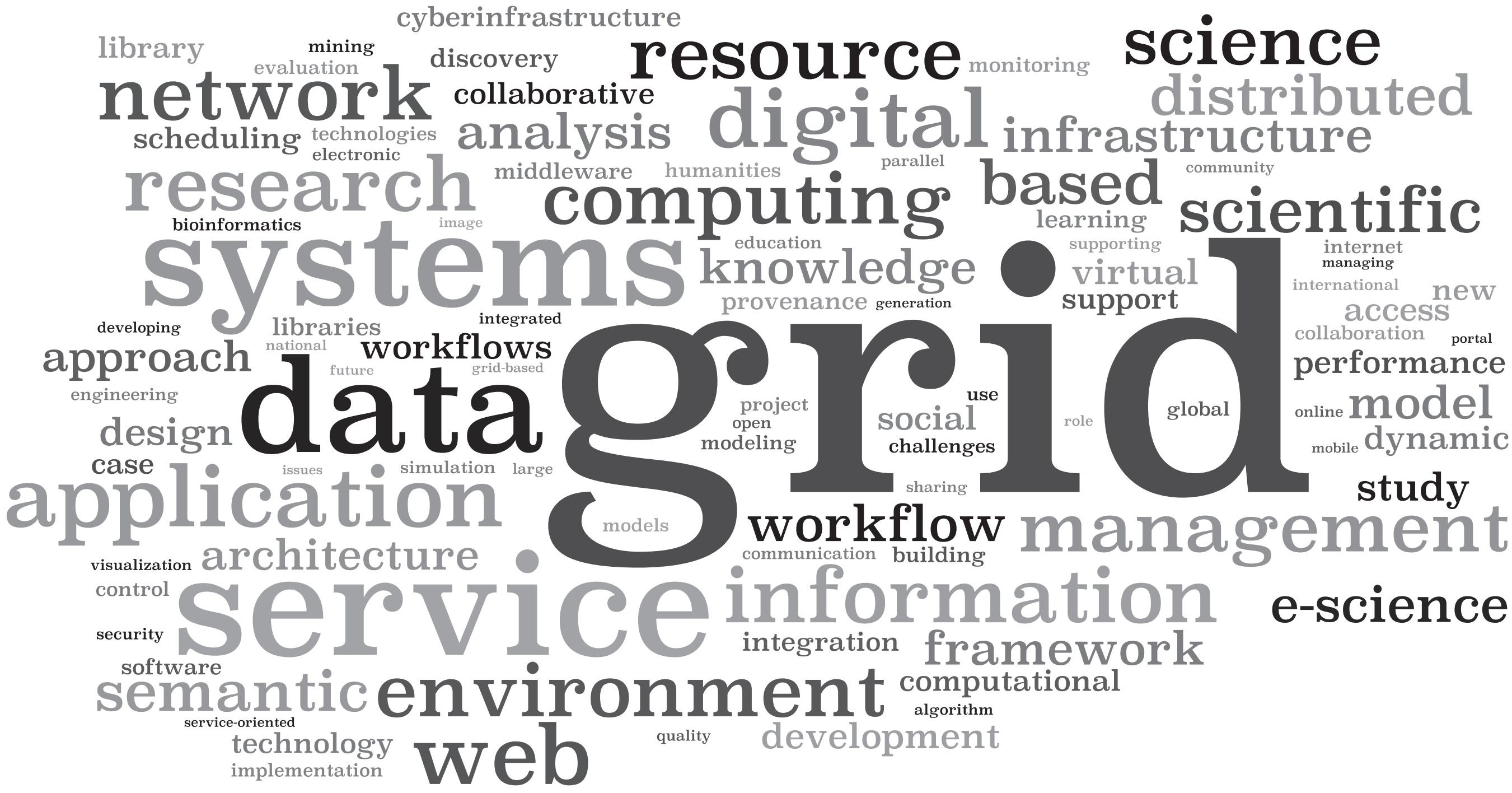

| Top 25 Scopus e-Research title word frequencies | |
|---|---|
| 1. grid | 1335 |
| 2. service | 560 |
| 3. systems | 515 |
| 4. data | 514 |
| 5. application | 331 |
| 6. web | 326 |
| 7. digital | 316 |
| 8. information | 302 |
| 9. research | 298 |
| 10. network | 291 |
| 11. environment | 290 |
| 12. resource | 285 |
| 13. computing | 282 |
| 14. management | 274 |
| 15. science | 264 |
| 16. scientific | 245 |
| 17. based | 239 |
| 18. semantic | 224 |
| 19. distributed | 217 |
| 20. e-science | 198 |
| 21. workflow | 196 |
| 22. analysis | 192 |
| 23. infrastructure | 189 |
| 24. knowledge | 185 |
| 25. framework | 182 |
| Total terms in sample ^ | 1312 |
| Total word instances ^ | 30210 |

| Top 25 NSF cyberinfrastructure title word frequencies | |
|---|---|
| 1. research | 74 |
| 2. cyberinfrastructure | 55 |
| 3. science | 53 |
| 4. collaborative | 50 |
| 5. network | 49 |
| 6. engineering | 44 |
| 7. workshop | 38 |
| 8. performance | 34 |
| 9. high | 31 |
| 10. ci-team | 27 |
| 11. data | 25 |
| 12. grid | 23 |
| 13. middleware | 23 |
| 14. computing | 21 |
| 15. demonstration | 19 |
| 16. development | 17 |
| 17. petascale | 17 |
| 18. university | 17 |
| 19. education | 16 |
| 20. computational | 15 |
| 21. connection | 15 |
| 22. project | 15 |
| 23. applications | 14 |
| 24. new | 14 |
| 25. services | 14 |
| Total terms in sample ^ | 94 |
| Total word instances ^ | 1280 |

^ Common words and terms with fewer than 5 instances removed

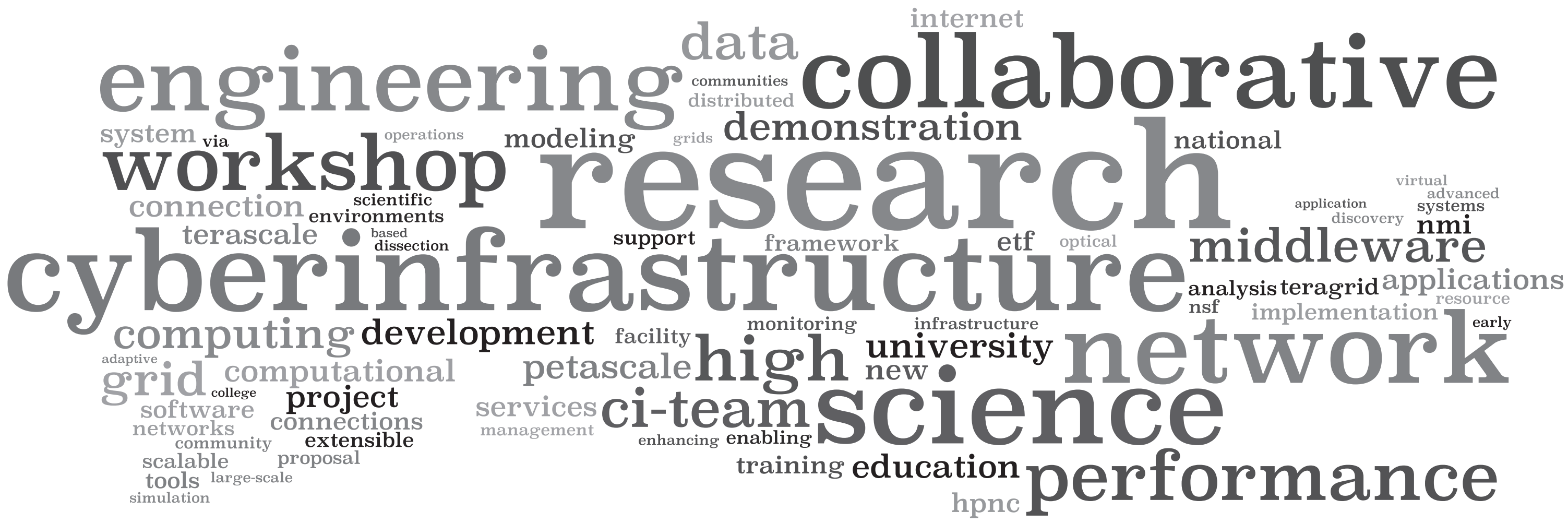

## Word cloud of NSF Office of Cyberinfrastructure award titles, 2000-2008

**Table 1**

**Table 1. Scopus e-Research papers, 1996-2008 (n=5206) compared to Wuchty et al. team size**

| Subject area | N[1] | Authors (mean) | Wuchty et al. (mean, 1996-2000) | Significance (p) | |
|---|---|---|---|---|---|
| Agriculture | 53 | 5.87 | 3.29 | 0.0286 | |
| Arts & Humanities | 43 | 1.44 | 1.10 | 0.0474 | |
| Biochemistry | 641 | 4.20 | 4.46 | 0.0850 | |
| **Business** | **199** | **2.52** | **1.66** | **<0.0001** | *** |
| Chemical Engineering | 26 | 3.54 | 2.74 | 0.0597 | |
| Chemistry | 51 | 4.82 | 3.69 | 0.0115 | |
| **Computer Science** | **2821** | **3.85** | **2.39** | **<0.0001** | *** |
| **Economics** | **57** | **2.25** | **1.71** | **0.0002** | ** |
| Energy | 25 | 4.56 | 2.64 | 0.1504 | |
| **Engineering** | **1634** | **3.65** | **2.94** | **<0.0001** | *** |
| Environmental Sciences | 123 | 3.17 | 2.98 | 0.4143 | |
| Immunology | 5 | 8.00 | 5.38 | 0.4246 | |
| Materials Sciences | 41 | 3.20 | 3.46 | 0.4508 | |
| **Mathematics** | **894** | **3.85** | **1.84** | **<0.0001** | *** |
| Medicine | 174 | 5.25 | 4.39 | 0.0126 | |
| Neurosciences | 13 | 5.08 | 4.05 | 0.2215 | |
| Nursing | 22 | 4.00 | 2.12 | 0.0156 | |
| Pharmacology | 22 | 4.18 | 4.31 | 0.8238 | |
| Physics | 191 | 4.41 | 4.05 | 0.1794 | |
| Psychology | 122 | 2.62 | 2.57 | 0.7582 | |

* p < .01, ** p < .001, *** p < .0001; Wuchty et al. data significance as difference from Wuchty mean in Scopus sample for fields where which have directly comparable data, as measured by one-sample t-test using SAS 9.1; Source Wuchty, Jones and Uzzi (2007).

1. Total of individual subject area classifications adds up to more than sample n due to multi-discipline papers being reported in more than one classification.